# HIV, TB and ART: the CD4 enigma[1]

Brian G. Williams* and Eleanor Gouws†

\* South African Centre for Epidemiological Modelling and Analysis (SACEMA), Stellenbosch, South Africa
† Joint United Nations Programme on AIDS (UNAIDS), Johannesburg, South Africa

Correspondence to BrianGerardWilliams@gmail.com

## Abstract

The concentration of $CD4^+$ T-lymphocytes (CD4 count) in a person's plasma is widely used to decide when to start HIV-positive people on anti-retroviral therapy (ART) and to predict the impact of ART on the future course of HIV and tuberculosis (TB). However, $CD4^+$ cell-counts vary widely within and among populations and depend on many factors besides HIV-infection. The way in which CD4 counts decline over the course of HIV infection is neither well understood nor widely agreed. We review what is known about CD4 counts in relation to HIV and TB and discuss areas in which more research is needed to build a consensus on how to interpret and use CD4 counts in clinical practice and to develop a better understanding of the dynamics and control of HIV and HIV-related TB.

## Introduction

$CD4^+$ T-lymphocytes (CD4 cells) play a key role in HIV-infection. HIV uses CD4 cells to replicate and resting CD4 cells can continue to harbour the virus even with effective ART.[1,2] During the acute phase of infection the virus infects CD4 cells, viraemia increases exponentially to very high levels and the concentration of CD4 cells in the plasma (CD4 count) drops by about 25%. After about one month cytotoxic T-cells control the primary infection, viraemia falls and CD4 counts recover. During the chronic phase of infection, which can last from about 2 to 20 years, the virus replicates in the peripheral lymphatic tissue and CD4 counts decline slowly[3-7] eventually to zero, but with increasing mortality[8] and incidence of opportunistic infections,[9] including tuberculosis.[7,10,11] The rate at which CD4 counts decline is determined by the rate of permanent viral replication. Once HIV-positive people start anti-retroviral therapy (ART), CD4 counts increase[12] and overall mortality[13] and the incidence of TB[14] decline.

Because of their association with disease progression, CD4 counts are used as a prognostic marker for deciding when to start treatment.[15-24] They have also been used to model the impact of starting ART at different CD4 counts[25-27] and for predicting the impact of ART on TB.[7,10]

The recommended CD4 count threshold, below which people should start ART, has varied greatly over the years,[28] and the recommendations have sometimes included a viral load threshold. The CD4 count thresholds are generally set at 200/µL, 350/µL, or 500/µL for no apparent reason[25-27] and do not take into account the substantial variation within and among populations. Here we argue for the need to develop a better understanding of the in-host dynamics of CD4 counts and the implications that this has for the management of HIV in individuals and the control of HIV in populations.

## Observations

### HIV drives TB

The most common opportunistic infection associated with HIV is TB and, especially in East and southern Africa, the epidemic of HIV has driven up the incidence of TB. As illustrated in Figure 1 the epidemic of HIV in Kisumu, Kenya started in the early 1980s and reached half of its peak value in 2000; the epidemic of TB followed with a delay of 5 years.[29] When the epidemic of HIV stabilized the incidence of TB had increased by 11 times.

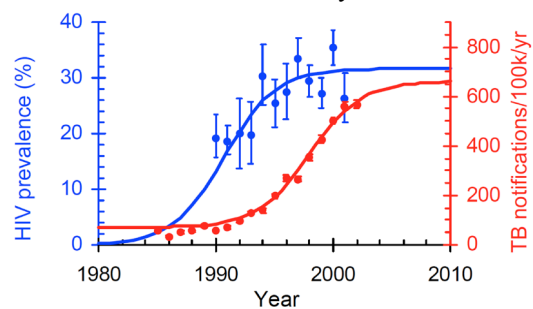

Figure 1. Epidemic trends for HIV and TB in Kisumu, Kenya. Blue: adult prevalence of HIV; Red: TB notifications.[29]

**The incidence rate ratio**

In order to understand the impact of HIV on TB we need to know the TB incidence rate ratio, *IRR*, which is the relative risk of TB in HIV-positive and HIV-negative people. The first and the most direct demonstration of the *IRR* was published by Corbett *et al.*[30] using data from gold miners in South Africa. In 1990 the prevalence of HIV among South African women attending public antenatal clinics (ANC) was 0.5%, by the year 2000 it had reached 24%.[31] During this time the *IRR* increased from 2 to 10 as shown in Figure 2. Of equal importance was the observation that the incidence of TB in HIV-negative mineworkers remained unchanged over the same period.

Other things being equal, the risk of TB infection is proportional to $P_{TB}$, the prevalence of people with active TB disease. As a first approximation $P_{TB}$ is equal to the incidence, $I_{TB}$, times the disease duration, $D_{TB}$, so that

$$P_{TB} = I_{TB} \times D_{TB}. \qquad 1$$

Now consider the period from 1991 to 1999 as shown in Figure 2. Since the incidence of TB in HIV-negative people did not change and we can assume that the disease duration in HIV-negative people did not change we can

---

1   This is a development of a presentation given by Brian Williams at the 'Modelling and Quantitative Research Priorities' meeting of the TB Modelling and Analysis Consortium (TB MAC), funded by the Bill and Melinda Gates Foundation, 28th and 29th September 2012, Johannesburg, South Africa conclude that the overall prevalence of TB did not change. However the incidence of TB in HIV-positive people did increase significantly and to maintain the same risk of infection in HIV-negative people, and therefore the same prevalence, there must have been a corresponding decrease in the disease duration in HIV-positive people.

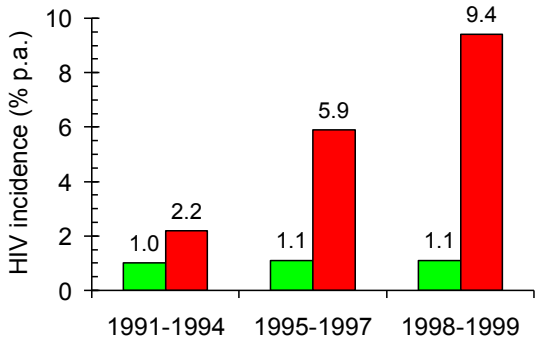

Figure 2. The incidence of TB among gold miners in South Africa between 1991 and 1999. Green: HIV-negative; red: HIV-positive.[30]

The average TB disease duration in HIV-negative people is 1 to 2 years[32] so that the disease duration in HIV-positive people, without anti-retroviral treatment, must be 1 to 2 months. From this it follows that even during a severe epidemic of HIV, conventional TB control should continue to work equally well for HIV-negative people. Since TB disease progresses so rapidly in HIV-positive people it will be very difficult to find HIV-positive people with active TB before they die. The control of HIV-related TB therefore depends largely on controlling HIV and ideally finding people with HIV and starting them on ART before they develop TB.

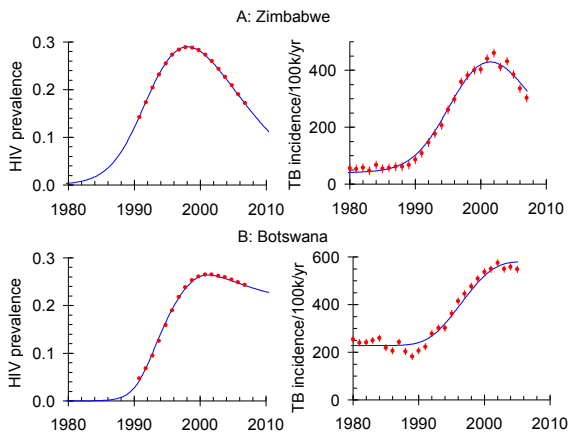

Figure 3. The epidemics of HIV and TB in A: Zimbabwe and B: Botswana. The *IRR* in Zimbabwe is 34 and in Botswana is 7.4.

*Variations in the incidence rate ratio*
Direct data on the *IRR* are hard to obtain and the Corbett study[30] is exceptional. However, we can estimate the *IRR* indirectly from time trends in the incidence of TB and the prevalence of HIV. Figure 3 shows data for Zimbabwe and Botswana.[10] While the HIV-epidemics, at least up to the peak, are similar in the two countries, there are three important points to be made. First, before the epidemic of HIV started, the incidence of TB in Zimbabwe was about five times lower than in Botswana. Second, the absolute increase, after the HIV epidemic peaked, was about 350 cases per 100k per year in both countries. Third the *proportional* increase in the incidence of TB in Zimbabwe was about three times greater than in Botswana.

To estimate the incidence rate ratio from the data in Figure 3 we let $I_{TB}$ be the incidence of TB, $I_{TB}^+$ and $I_{TB}^-$ in HIV positive and negative people, respectively, and $P_{HIV}$ be the prevalence of HIV.† Then

$$I_{TB} = P_{HIV} \times I_{TB}^+ + (1 - P_{HIV}) I_{TB}^-$$
$$= [P_{HIV} \times IRR + (1 - P_{HIV})] I_{TB}^- \qquad 2$$

so that

$$IRR = \frac{R - (1 - P_{HIV})}{P_{HIV}} \qquad 3$$

where $R = I_{TB}/I_{TB}^-$ is the ratio of the TB incidence at a later time to the incidence before the HIV epidemic started.‡ From Equation 3 and the data in Figure 3, the *IRR* in Zimbabwe is 34, almost five times higher than in Botswana where it is 7.4. We consider possible explanations below.

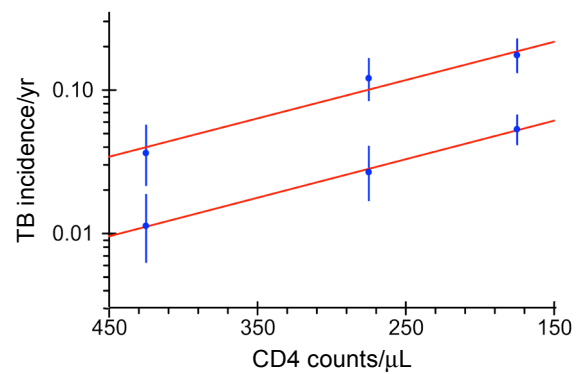

Figure 4. TB incidence as a function of CD4 counts in HIV-positive people from studies in public hospitals in Italy[33] and in Cape Town.[34] For every CD4 count drop of 100 cells/μL the incidence of TB increases by 62%.

**Correlates of TB incidence**

*CD4 counts*
CD4 counts are significantly correlated with the incidence of TB in HIV-positive people (Figure 4). While correlation does not imply causation, the linear decline of CD4 counts after infection with HIV[35] will lead to an exponential rise in the incidence of TB,[33,36] people with advanced HIV infection will be more likely to develop TB, and this can account for the five year delay in the rise of TB.[10] Furthermore, this provides an explanation for the observation that the average value of the CD4 count in HIV-positive people presenting with TB is about 200/μL.[7]

An association with CD4 counts can also explain the 3.8 (1.6−15.2) fold increase in the incidence of TB during the acute phase of HIV-infection.[33] During the acute phase CD4 counts fall by about 25% (9%−41%).[10,35] If we assume that the CD4 count in HIV-negative people is about 1,000 cells/μL then a drop of 250 cells/μL would increase the incidence of TB by a factor of about 3.

---

† This result refers to an average over all HIV positive people in different stages of disease progression at a given time.
‡ The incidence rate ratio can also be calculated from the prevalence of HIV in TB patients and the general population.[61]

*Obesity*

There is a remarkably strong log-linear relationship between the incidence of TB and body mass index (BMI) in HIV negative people; the incidence of TB falls by 13.8% (± 0.4%) across four orders of magnitude in the incidence of TB and BMI values ranging from 17 to 32kg/m$^2$, and in subjects ranging from residents of Hong Kong pensioners, to Finish smokers and US Navy recruits.[37]

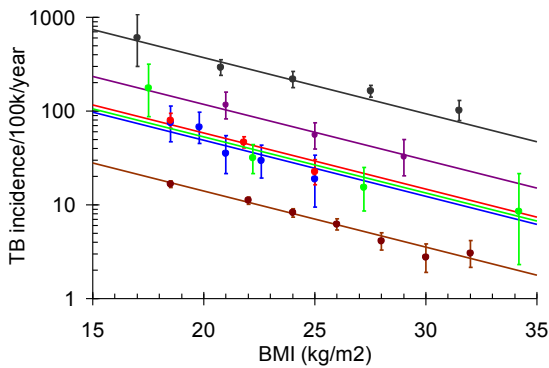

Figure 5. Log-linear relationship between BMI and the incidence of TB across nearly four orders of magnitude and BMI ranging from 17 to 32kg/m$^2$. The incidence of TB falls by 13.8% (±0.4%) per unit increase in BMI.[37] Black: Hong Kong health centres for the elderly;[38] purple: male smokers in a Finnish cancer trial;[39] red: US Navy recruits;[40] green, NHANES, USA (P. Cegielski, pers. comm.); blue, US Navy recruits;[41] brown: Mass Radiography Survey, Norway[42] ($p = 10^{-51}$).

Since HIV is a wasting disease it is possible that the association between CD4 counts and the incidence of TB is partly dependent on a progressive decline in BMI with time since HIV infection so that BMI could confound the apparent relationship between CD4 counts and TB. To achieve a three fold increase in TB as a result of a decline in BMI, people would have to lose about 8 kg/m$^2$ of BMI during the acute phase; to bring about a ten-fold increase in the incidence of TB in people with an established HIV infection they would have to lose about 16 kg/m$^2$ of BMI neither of which seem plausible. While few studies seem to have collected time series data on BMI over the course of an HIV-infection at least one study suggests that BMI only falls significantly during the last six months before the onset of AIDS.[43] One of the few studies on BMI and CD4 counts[44] suggests that increasing BMI is associated with increasing CD4 cell counts but the effect would only explain a small part of the decrease in TB with increasing CD4 cell counts. In spite of the very strong association between BMI and TB this seems unlikely to confound the relationship between CD4 counts and TB.

## CD4 count variation among populations

CD4 count distributions vary widely among different populations.[35] As shown in Figure 6 four studies in HIV-negative people in Ethiopia[45,46] (red 1) consistently give mean values of about 750/μL with a 95% range of about ±200/μL. However the CD4 counts vary greatly both in their ranges, from ±90/μL in Tanzania[47,48] to ±530/μL in Botswana[49] (blue 2), and in their means, from 1,150/μL[50] in Uganda to 591/μL in Botswana[49] (green 3).

CD4 cell counts are highly variable among and within individuals depending on their sex, stress levels, fitness levels, time of day, pregnancy, age and many other things.[51] If CD4 counts are to help us to understand the dynamics of HIV infection and the incidence of opportunistic infections as well as to provide a useful prognostic marker, all of this variation needs to be allowed for and, if possible, understood.

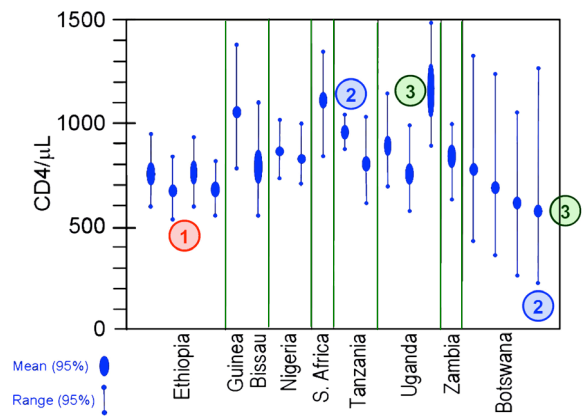

Figure 6. CD4 count variability in HIV negative populations. The central ellipses indicate 95% confidence limits on the estimated mean values and the points at the end of each line indicate 95% estimates of the population distribution. Circled numbers are discussed in the text.

## Modelling CD4 count progression

In order to understand the relationship between CD4 counts and disease progression, individual prognosis and the incidence of opportunistic infections, it will be necessary to model the in-host dynamics of CD4 counts in relation to HIV infection. First, this must include the variation in CD4 counts within a given population (Figure 6). Second it must include the variation in the mean and the spread of CD4 counts among populations (Figure 6). Third, it must include the variation in survival in people infected with HIV which ranges from 2 year to 20 years.[52] Fourth, it must allow for the age of the infected person since the median survival after infection with HIV varies from 16 years in those infected (horizontally) before the age of 5 years to 4 years in those infected after the age of 65 years.[52] Fifth, it must allow for the state of the epidemic; when the prevalence is rising people are more likely to have been infected recently and when it is falling they are less likely to have been infected recently.

A simple phenomenological model was developed to allow for the variation in initial CD4 counts and survival, taking into account the distribution of CD4 counts in the relevant population, and assuming that survival is independent of the initial CD4 count.[35] With this model it has been possible to predict the CD4 count distribution in HIV-positive people from the distribution in HIV-negative people,[35] the variability in the rate of decline of CD4 counts,[53] and the one-year mortality of post-partum women,[8] in all cases with no free parameters apart from a normalization factor. The assumption of a linear decline in CD4 counts with time since infection is consistent with at least one detailed model of the immune response to HIV[54] but much further work is needed in this regard.

## CD4 recovery under ART

Modern anti-retroviral therapy can suppress viral load in people infected with HIV by up to 10 thousand times[55-57]

and this can be sustained for at least seven years[55,56] and probably indefinitely. It also leads to significant increases in CD4 counts as shown in Figure 7.

What is remarkable about the data in Figure 7 is the consistency of the rate of increase and the asymptotic values with the only difference being the CD4 count at the start of treatment. Given the great variability in CD4 cell counts in HIV-negative people this consistency in the recovery rates in HIV-positive people is unexpected. The CD4 counts all converge exponentially at a rate of 224/µL/year (±35/µL/yr) to an asymptote that is 371/µL (±16/µL) above the value at the start of treatment. This suggests that if ART is started late, the infected person's immune function will never fully recover. It would be of interest to develop a simple model of CD4 cell recovery under ART that can help to explain the data in Figure 7.

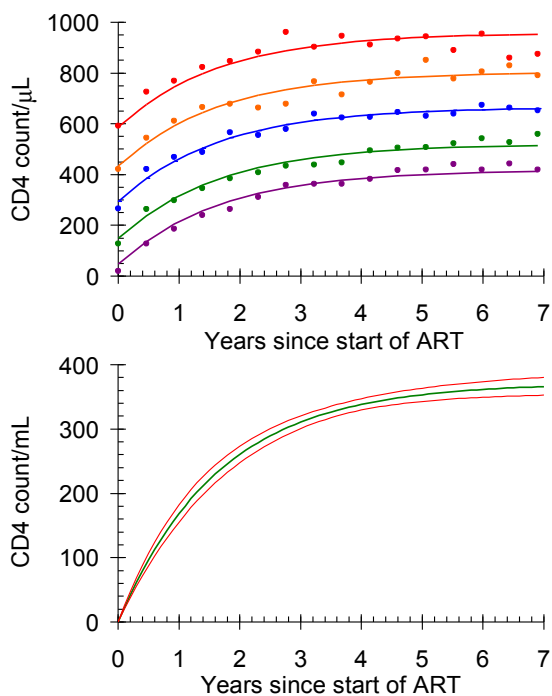

Figure 7 A. CD4 count recovery under ART. The fitted lines differ only in their initial starting points and have the same rate of increase and the same asymptote. B. The fitted curve and confidence bands with data offset to their initial values. Asymptote: 371 (355−389)/µL; rate of increase: 224 (192−261)/µL. Data from Gras *et al.*[58] but see also De Beaudrap *et al.*[59]

There is good data to suggest that the 'recovery' of the immune system with respect to TB is never complete as shown in Figure 8 where even those people whose CD4 count rose to 800/µL still suffered a TB incidence about 4 times the rate in HIV-negative people.

Several studies have shown that ART reduces the incidence of TB by about 60% irrespective of the CD4 count at which treatment starts.[10,60] This apparent consistency in the proportional decline in the incidence of TB is consistent with the observation that the increase in CD4 counts after the start of ART is independent of the CD4 counts at the start of treatment (Figure 7) and that the incidence of TB varies exponentially with the change in CD4 counts (Figure 4). In the studies showing a consistent decrease in TB incidence the mean duration of follow up was between 1 and 2 years during which time the increase in CD4 counts must have been between 200 and 250 cells/µL. The estimate in Figure 4 of a 62% increase in TB for every 100 cells/µL decline in CD4 counts would suggest that ART should reduce the incidence of TB after 1 to 2 years by between 70% and 80%, similar to the 60% decline noted above.

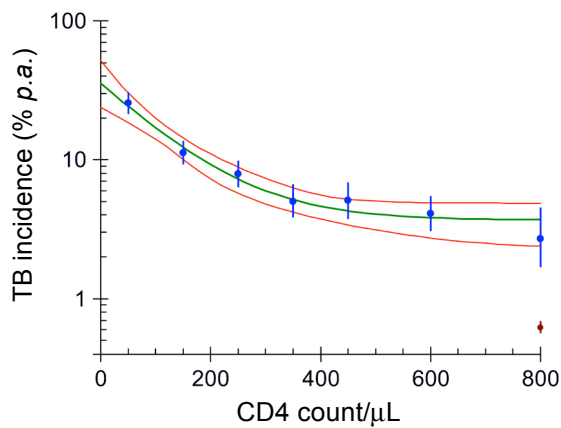

Figure 8. Decline in the incidence of TB among people on ART as a function of CD4 count. Brown point: HIV-negative people in the same community.[14]

**Incidence rate ratios revisited**

Two explanations have been proffered for the variability in the *IRR*. The first follows from the fact that we are comparing the prevalence of HIV with the incidence of TB so that clustering of the two infections in the same sub-populations could produce the observed effect. For example, suppose that the incidence of TB in those infected with HIV increased by a fixed factor. Then if one added to the population an equal number of people without TB or HIV the prevalence of HIV would be halved, the incidence of TB per person would also be halved, before and after the introduction of HIV, but the *proportional* increase in TB would remain the same and the *IRR* would appear to have doubled. This argument is supported by the observation that in concentrated epidemics in developed countries the *IRR* appears to be about 60, very much higher than in generalized epidemics in most developing countries.[61] The putative explanation for this is that in developed countries the same subset of often marginalized people is at high risk of both HIV and TB. An alternative explanation is that in countries like Botswana where the CD4 counts in HIV-negative people are very low they are more susceptible to developing TB when they are HIV-negative while in Zimbabwe, where the CD4 count in HIV-negative people is much higher, they are less susceptible to developing TB when they are HIV-negative. But because people in Botswana start with a low CD4 count while people in Zimbabwe start with a high CD4 count, the CD4 decline in Zimbabwe is greater than in Botswana and the increase in TB is correspondingly greater in Zimbabwe.

**Discussion**

**What do we know?**

Certain aspects of the HIV-TB epidemic are clear: the epidemic of HIV drives up the *incidence* of TB with a delay of about 5 years; the incidence rate ratio is substantial but varies among African populations from about 5 to about 20; HIV does not appear to increase the *prevalence* of TB; ART reduces the incidence of TB but only by about 60% and apparently by the same amount at all CD4 counts. CD4 counts vary greatly both within and among populations and even within individuals over time.

**What do we understand?**

HIV does not affect the incidence of TB in HIV negative people, and therefore the overall prevalence of TB, because the increase in the incidence of disease is balanced by a decrease in the duration of disease and to some extent by the fact that HIV-positive people are more likely to develop disseminated TB and therefore to be less infectious than HIV-negative people.

**What can we tentatively conclude?**

Conventional TB control among HIV-negative people should work equally well in the face of an epidemic of HIV. CD4 counts may provide the link between HIV and the increasing incidence of TB but this is hard to establish definitively. Since the rate and extent of CD4 recovery is independent of the CD4 count at the start of ART the proportional reduction in the incidence of TB is independent of the CD4 count at the start of treatment and is never complete.

**What do we still not understand?**

Given that CD4 cell counts are so variable and the fact that HIV relies on CD4 cells to replicate, what are the consequences of this for the progression of HIV infection in individual people? Are the CD4 counts in the peripheral lymphatic tissue relatively constant even though the CD4 counts in the plasma fall over the course of infection? Why does it appear to be the case that survival is independent of the CD4 cell count before infection? If CD4 counts are predictive of the risk of TB in HIV-positive people, is this also true in HIV negative people? Since the decline in CD4 cell counts is so variable when people are infected with HIV, why is the recovery so consistent? Why does the incidence rate ratio vary by a factor of up to 5 among different but seemingly similar populations and epidemics of HIV? Do CD4 cell counts have any prognostic value for individual patients? Can we realistically model the immune response to TB and learn important lessons about how to control TB in the presence of HIV?